# Spatial variation of short-range order in amorphous intergranular complexions

Zhiliang Pan[a], Timothy J. Rupert[a,b,*]

[a] Department of Mechanical and Aerospace Engineering, University of California, Irvine, California 92697, USA
[b] Department of Chemical Engineering and Materials Science, University of California, Irvine, California 92697, USA
*To whom correspondence should be addressed: trupert@uci.edu

## Abstract

Amorphous materials lack long-range order but short-range order can still persist through the recurrence of similar local packing motifs. While the short-range order in bulk amorphous phases has been well studied and identified as an intrinsic factor determining the material properties, these features have not been studied in disordered intergranular complexions. In this work, the short-range order in two types of amorphous complexions is studied with a Voronoi tessellation method. Amorphous complexions can have three distinct regions: amorphous-crystalline interfaces, regions deep inside the films that have short-range order identical to a bulk amorphous phase, and transition regions that connect the first two regions. However, thin amorphous films contain only the amorphous-crystalline interface and the transition region, providing further evidence of the constraints imposed by the abutting crystals. The thickness of the transition region depends on film thickness at low temperatures but becomes thickness-independent at high temperatures. Similarly, the complexion short-range order is dependent on the interfacing crystal plane at low temperatures, but this effect is lost at high temperatures. Our findings show that amorphous complexions contain spatial gradients in short-range order, meaning they are both structurally and chemically different from bulk metallic glasses.

# Introduction

Complexions are interfacial states that exhibit phase-like behavior under different thermodynamic conditions [1]. One way of categorizing complexions is by their thickness or levels of adsorbate, where six types have been identified by Dillon et al. [2]: (I) a single layer of dopants, (II) clean grain boundaries, (III) bilayers, (IV) multilayers, (V) intergranular films of nanoscale equilibrium thickness, and (VI) wetting films. While complexions I–IV are often structurally ordered with legible layers of dopants, complexions V and VI are often disordered [3], in which case they can be broadly grouped as amorphous intergranular films. Complexion V is an amorphous region sandwiched between two amorphous-crystalline interfaces separated by a distance of a few nanometers, with the thickness determined by interfacial thermodynamics, and is thus called a nanoscale intergranular film. Complexion VI is in fact a bulk amorphous phase sandwiched between two amorphous-crystalline interfaces. Amorphous intergranular films can lead to activated sintering [4-6] in ceramics and refractory metals [7, 8] and abnormal grain growth in other materials [2, 9-11]. They can also assist grain boundary sliding [12, 13] and suppress crack formation induced by grain boundary-dislocation interactions at grain boundaries [14, 15], making it possible to fabricate tougher nanocrystalline materials [16]. Since amorphous complexions are important features for material fabrication, design, and optimization, the detailed atomic structure from which the physical and mechanical properties originate [17] are of great interest.

While long-range order determines the deformation mechanism of crystalline materials, short-range order from the repetition of local packing motifs controls the plastic deformation of metallic glasses [18]. It has been observed that full-icosahedral clusters, a common five-fold symmetric, quasi-crystalline packing motif in metallic glass, can increase strength but deteriorate ductility [19, 20]. Shear banding, one of the possible plastic deformation mode common in bulk metallic glasses, has been connected to the breakdown of full icosahedral clusters from the microscopic perspective [21] and a decrease in the fraction of perfect icosahedral clusters is observed during the plastic deformation [22]. The short-range order in a bulk amorphous phase can be affected by fabrication parameters as well. For example, decreasing cooling rate can increase the fraction of icosahedral clusters [23]. Short-range order can also be affected by ion irradiation which breaks stiff icosahedral packing motifs, leading to a softening effect within the cascade region [24]. In addition, short-range order has been connected with the thermal stability [25] and magnetic properties of metallic glasses [26]. Short-range order is also a theoretical foundation of an efficient cluster packing model [27], which has successfully predicted or estimated the density [28], glass forming ability, and thermal stability [29, 30] of metallic glasses. Since short-range order plays such an important role in determining the properties of bulk amorphous phases, the local structural order of amorphous complexions is also of interest.

Experimental observations [31, 32] have shown the repeated distribution of the same type of packing clusters in nanoscale intergranular films at the vicinity of abutting crystals, indicating the existence of short-range order at the amorphous-crystalline interfaces. However, these studies concentrated only on the structural order at amorphous-crystalline interfaces. Whether the short-range order exists and what type of structural order dominates the disordered interior was not investigated due to the limited resolution of experimental techniques. Nanoscale intergranular films and wetting films are both described as being structurally disordered, but are two different types of structures and have different alloy compositions, with the solute composition in wetting films higher than that in nanoscale films [3]. While the short-range order in bulk amorphous phase has been shown to be affected by alloy compositions [3], whether and how the difference in alloy composition in the two types of complexions leads to different structural order remains unknown. The alloy composition also changes with thickness for nanoscale intergranular films [3], suggesting that the short-range order might change with the film thickness as well. The short-range order observed in amorphous complexions is believed to be imposed by abutting crystals, decaying into the film interior with a distance from the amorphous-crystalline interface [33-35]. The test of this hypothesis makes it highly necessary to check spatial variation of structural order along the film thickness direction. Finally, the role of abutting crystals also implies that the structural order imposed by different crystals planes should be different. While atomistic simulations have the advantage of precisely describing atomic level microstructures over experimental observations and thermodynamic models, the vast majority of studies to date [34-39] have concentrated on film structure in ceramic systems. Much less work (see, e.g., [17]) has been performed for complexions in metallic systems.

In this work, the short-range order of nanoscale and wetting amorphous intergranular films in Cu-Zr binary alloys is analyzed with a Voronoi tessellation method [40, 41]. This analysis method can provide detailed information about local clustering and has been successfully used in prior studies to describe bulk amorphous phases. Three distinct regions of structural order are identified, with the amorphous-crystalline interface, a transition region, and in the region deep within thick wetting films all having unique short-range order signatures. The transition region depends on overall film thickness at low temperature, but this effect goes away at high temperature. However, the structural order in amorphous-crystalline interfaces is always independent of complexion type and overall film thickness. Complexion short-range order is also shown to depend on the interfacing crystal plane at low temperature, but again this effect does not exist at high temperatures. This finding suggests that the constraint of the abutting crystals is reduced as temperature increases.

**Methods**

Complexion formation was simulated with hybrid atomistic Monte Carlo/molecular dynamics simulations in Cu-Zr binary alloys at 600 K and 1200 K using the Large-scale Atomic/Molecular Massively Parallel Simulator (LAMMPS) code [42], with 1 fs integration time step used for all molecular dynamics simulations. Embedded-atom method potentials were used to describe the Cu-Cu and Zr-Zr interactions, while a Finnis-Sinclair potential was used to describe the Cu-Zr interactions [43]. The simulations started from a Cu bicrystal which was first equilibrated using a Nose-Hoover thermos/barostat at zero pressure and then doped with Zr solutes. The doping process was simulated with a Monte Carlo method in a variance-constrained semigrand canonical ensemble [44] while structural evolution and relaxation was simulated with molecular dynamics, with additional simulation details provided in reference [3]. This hybrid simulation technique can model dopant segregation and complexion transitions at different grain boundaries under a variety of thermodynamic conditions. Adaptive common neighbor analysis [45] was used to identify the local crystal structure of each atom, with face-centered cubic atoms colored green, hexagonal close packed atoms red, body-centered cubic atoms purple, icosahedral atoms yellow, and other atoms white. Zr atoms were highlighted as dopants with blue color. Figure 1(a) shows a bicrystal Cu sample containing two $\Sigma 5$ (310) boundaries. Samples containing $\Sigma 11$ (113), twist (100) 36.86°, twist (100) 10.39°, and twist (111) 30° boundaries were also studied since connecting with different crystal planes of the abutting grains is expected to affect short-range order [33]. To facilitate a comparison, bulk amorphous phases were also produced through the hybrid simulation technique in a periodic simulation box having dimensions of ~8×8×8 nm. The Zr composition is fixed to 25.8 at. % at 600 K and 11.5 at. % at 1200 K, which are the Zr compositions of wetting films obtained at the two temperatures.

The samples were quenched with a conjugate gradient energy minimization technique to remove thermal noise yet preserve the grain boundary structure. A Voronoi tessellation method was then used to analyze the Voronoi polyhedron associated with each atom, which is enclosed by all of the bisecting planes of the lines connecting the center atom and its neighboring atoms. The Cu/Zr atomic size ratio is set to be 0.804 [27] to weight the bisection. Voronoi polyhedra can be described using the index notation $<n_3,n_4,n_5,n_6,...>$, where $n_i$ denotes the number of Voronoi polyhedron faces with $i$ edges and the sum of $n_i$ is the coordination number of the centered atom, since each face is associated with a neighbor of the center atom. The short-range order of a material then can be defined as the recurrence of the same type of polyhedron. The minimum edge length on the polyhedron is set to be 0.6 Å to make sure that the polyhedra associated with atoms in the grain interior are identified as <0,12,0,0>, a building block for perfect face-centered cubic crystals. To isolate the amorphous intergranular films, face-centered cubic atoms with <0,12,0,0> polyhedron are excluded when calculating radial distribution functions and performing statistical analysis of the complexion structure.

## Results and Discussion

With increasing Zr content, the grain boundaries either transform gradually into nanoscale films and then wetting films, as shown in Figure 1(b), which gives the grain boundary composition and grain boundary thickness as a function of global composition, as well as the typical equilibrium complexion structures found at these different stages. One way to show the structural order of a material is through its radial distribution function, which gives the number density of particles in a spherical shell centered at a reference particle. Figure 2(a) shows such a function for the bulk amorphous phase obtained at 600 K. The peaks of this function vanish rapidly with increasing pair separation distance, indicating the lack of long range order. Since short-range order is related to the first peak and the free surfaces of isolated amorphous films introduce non-negligible error at long pair separation distances, the comparisons between amorphous intergranular films and bulk amorphous phase are made only at short pair separation distances, as shown in Figure 2(b). The locations of the first peaks of different samples match almost perfectly, except that the function values of the complexions at the first peak and first valley (inset to Figure 2(b)) are higher than that of the bulk phase. Realizing that amorphous complexions have amorphous-crystalline interfaces while a bulk phase does not, a radial distribution function of the interior of a 5.30 nm thick film (cut to isolate the inner 4 nm thick region) is shown together with the bulk amorphous phase in Figure 2(c). The differences at the first peak and the valley after disappear, indicating that the structural order at amorphous-crystalline interfaces is different from that in film interior.

Structural analysis based on radial distribution functions only gives a rough estimation of the difference and similarity of structural order between different amorphous complexions and the bulk amorphous phase. Much more detailed information can be obtained by analyzing the building blocks of short-range order, or the Voronoi polyhedron associated with each atom. Figure 3 shows the planar density (number of atoms per unit cross-sectional area) of Cu atoms associated with the collection of the 10 most frequent polyhedra in the amorphous films created at a $\Sigma 5$ (310) grain boundary at 600 K, with the data plotted as a function of film thickness. It should be noted that film composition is only constant for wetting films, as grain boundary composition increases with film thickness for nanoscale films, as shown in Figure 1(b). Regardless of the grain boundary composition, the planar density of each polyhedron increases with the film thickness in a roughly linear fashion, but with different starting values and slopes. Atoms associated with <0,10,2,0> polyhedron have the highest density in the 1.85 nm thick film, but the density only increases slowly with film thickness and ends up as the fourth most frequent in the 5.30 nm thick amorphous film. In contrast, atoms associated with the <0,2,10,0> polyhedron are the third most frequent in the 1.85 nm thick film, but the density increases quickly with film thickness and ends up as the most common local order type in the 5.30 nm thick amorphous film. The inset to Figure 3 shows that linear fits to the <0,10,2,0> and <0,8,4,0> polyhedral types intercept with the vertical axis at ~9 and ~4 atoms/nm$^2$, respectively, whereas the fitted lines of <0,2,10,0> and <0,0,12,0> intercept with the horizontal axis at ~0.8 nm. This indicates that the former two types of polyhedra still exist when the film thickness converges towards zero, whereas the latter two types of polyhedra disappear at a film thickness of ~0.8 nm.

An amorphous intergranular film contains one disordered interior and two amorphous-crystalline interfaces that sandwich this interior. While the interfaces should always exist regardless of the film thickness, the disordered interior disappears when the film is thin enough. This suggests that the polyhedra that still exist when film thickness reaches zero are distributed at the amorphous-crystalline interfaces, while the polyhedra that have vanished are associated primarily with the film interior. To verify this hypothesis, the spatial distributions of the <0,10,2,0> and <0,2,10,0> polyhedra along the thickness direction of an amorphous film are given in Figure 4(a) and Figure 4(b), respectively. The positions of the two peaks in Figure 4(a) show that <0,10,2,0> polyhedra are most pronounced at the amorphous-crystalline interfaces. It should be noted that polyhedra that dominate amorphous-crystalline interfaces may also appear inside of the film interior, which explains why the planar density of atoms associated with the <0,10,2,0> polyhedron increases with increasing thickness, albeit at a very slow rate. However, the <0,2,10,0> polyhedron only appears inside the film interior, as confirmed by its exclusive distribution between the two dotted blue lines in Figure 4(b) that mark the boundaries of the disordered interior. The average thickness of amorphous-crystalline interfaces can be estimated by interpolating backward to find the film thickness at which the planar density of polyhedra that only appear in the film interior reaches zero, giving a value of ~0.8 nm. Since there are two amorphous-crystalline interfaces, the estimated thickness of the interfacial region should then be ~0.4 nm. With this information, the two amorphous-crystalline interfaces and interior of an amorphous film can be isolated and analyzed separately.

Figure 5(a) shows the distribution of a collection of the 10 most frequent polyhedra in amorphous-crystalline interfaces for the interfacial films shown in Figure 3. The fractions of each type of examined polyhedron near the interfaces are about the same, indicating that the amorphous-crystalline interfaces of both nanoscale intergranular films and wetting films have consistent short-range order, with <0,10,2,0>, <0,8,4,0>, <0,7,4,1>, and <0,9,2,1> polyhedra being the dominate contributions. Figure 5(b) shows the distribution of a collection of 10 most frequent polyhedra in the bulk amorphous phase and film interior. The fractions of each type of examined polyhedron in film interior are about the same as well, indicating that the interior short-range order of both nanoscale intergranular films and wetting films is not thickness dependent, with <0,2,10,0>, <0,0,12,0>, <0,4,8,0>, and <0,6,6,0> being the dominate polyhedra. However, the short-range order of the film interior is slightly different from the bulk amorphous phase, with the fraction of <0,2,10,0> and <0,0,12,0> polyhedra in the bulk phase being ~25% higher. However, this difference disappears, as shown in Figure 5(c), when only the very center of these films (the inner 2 nm of the two thickest wetting films) is considered. This observation indicates that the short-range order of the innermost region is fully consistent with that of bulk amorphous phase, while the short-range order of the region between the amorphous-crystalline interface and the innermost interior is different from a bulk phase.

It is therefore necessary to examine the structural order difference in detail. To this end, the fraction distribution of dominating polyhedra along film thickness direction is calculated, with Figure 6(a) showing the fraction distribution of <0,2,10,0> polyhedron from film boundaries to film centers. Here the fraction of one type of polyhedron at a position is calculated as that inside of a 0.4 nm thick bin centered at this position. The fraction is close to zero at the film boundary and increases first slowly when moving toward the edge of the amorphous-crystalline interface region (0–0.4 nm). Beyond this region, the fraction increases more quickly and finally reaches the level at bulk amorphous phases at distances that depend on film thickness. After that, the fraction fluctuates around that level. This indicates that, in addition to the amorphous-crystalline interface region and fully saturated region in the innermost film interior, there is a transition region in between. It should be noted that the rate of increase for different amorphous films is close at the amorphous-crystalline interfaces, but becomes divergent in the transition region. As a result, the fraction of <0,2,10,0> polyhedron in different films reaches the saturated value at different distances from the edge of the crystal. One exception is the 1.85 nm thick film, which does not reach this saturation level even at the film center (where the black curve ends). Figure 6(b) gives the thickness of the transition region as a function of thickness, measured as the distance between amorphous-crystalline interface and the position where the fraction of <0,2,10,0> polyhedron reaches 90% of the fraction (0.1) representative of a bulk amorphous phase. As shown by the black trend line, the transition region thickness increases slowly with film thickness. The thickness of the saturated region of different amorphous films is also plotted in Figure 6(b) and increases with increasing overall film thickness as well. The red trend line intercepts with the horizontal axis at a film thickness of ~1.8 nm, indicating that no saturated region can exist in amorphous films thin enough. This can be confirmed by the fraction distribution of the 1.85 nm film shown as a black curve in Figure 6(a).

The fraction distributions of other polyhedra (not shown here) also show similar trends, albeit with the exact transition region thickness depending on the polyhedron type. These findings suggest that the abutting crystals do have an influence on the short-range order of amorphous intergranular films, up to a certain point. The influence decays with the distance from the amorphous-crystalline interface and the structural order of the disordered complexion interior becomes progressively more similar to that of a bulk amorphous phase. Based on the spatial variation of the structural order, two possible configurations can be proposed for amorphous intergranular complexions. For a thick film, the configuration shown in Figure 7(a) occurs, with one fully saturated region sandwiched in between two transition regions which are in turn sandwiched by two amorphous-crystalline interfaces. However, for a thin film like that shown in Figure 7(b), only the transition region exists between the two amorphous-crystalline interfaces and no region in the film interior has structural order that is exactly like that of bulk amorphous phases. The two proposed configurations can help explain the difference between the two types of amorphous complexions. Wetting films are chemically and structurally identical to a bulk amorphous phase and the two amorphous-crystalline interfaces and transition regions are just necessary bridges for the structural order to transform from the fully saturated region to the two abutting crystals. For nanoscale films, the chemical and structural building blocks are different from that of a bulk phase, so the constraints of the crystals must be felt to stabilize the amorphous film. Figure 8 shows the fraction distribution of <0,2,10,0> polyhedron in amorphous films transformed from the Σ5 (310) grain boundary at 1200 K. The fraction increases with distance from the crystal in a manner that is similar with that shown in Figure 6(a). However, the rate of change of the <0,2,10,0> polyhedron fraction does not appear to depend on overall film thickness in this case.

Both our observations here and the literature [33-35] suggest that a certain level of structural order at the amorphous-crystalline interfaces is imposed by the abutting crystals, which in turn induce order that extends into the film interior. It is therefore expected that different crystal planes might impose different types and amounts of structural order on the amorphous-crystalline interfaces and transition regions of amorphous intergranular films. To test this hypothesis, the short-range order in amorphous films transformed from two other grain boundaries at 600 K is examined and compared with the structural order in films transformed from the Σ5 (310) boundary at this same temperature. Here, the crystal plane that is located at the grain boundary is used to label the grain boundary. For example, the interfacing plane is (310) for the amorphous films transformed from the Σ5 (310) boundary. The amorphous films facing (310) and (113) crystal planes have quite similar short-range order both at amorphous-crystalline interfaces and in film interior, as illustrated in Figure 9, where the planar density of a dominant polyhedron (<0,10,2,0>) at amorphous-crystalline interfaces and that in film interior (<0,2,10,0>) is plotted as a function of film thickness. However, the planar density of <0,2,10,0> polyhedron in films facing a (111) plane is always higher than that in the amorphous films facing the other two crystal planes, indicating that interfacing crystal planes do affect the short-range order in amorphous films.

While some dependence on the abutting crystals exists, it is quite possible that this influence is temperature dependent. Figure 10(a) shows the planar density of <0,10,2,0> and <0,2,10,0>

polyhedra as a function of film thickness in amorphous films transformed from various grain boundaries at 1200 K. In this case, the amorphous films facing different crystal planes have the same short-range order both at amorphous-crystalline interfaces and in film interior, meaning that the dependence of local structure on the abutting crystal plane is no longer present. It should be noted that the trend line of the planar density of <0,2,10,0> intercepts with horizontal axis also at ~0.8 nm. This indicates that the thickness of amorphous-crystalline interface at 1200 K is about the same as that at 600 K (~0.4 nm in both cases). The fraction distribution of the <0,2,10,0> polyhedron in ~10 nm thick films transformed from different grain boundaries at 1200 K is presented in Figure 10(b), demonstrating that films interfacing with different crystal planes also possess similar transition regions. The variety of short-range order at 600 K and uniform structural order at 1200 K shows that increasing temperature can reduce the effect of interfacing crystal planes on the structural order of amorphous complexions.

## Conclusions

In this work, the short-range order of amorphous intergranular complexions was analyzed with a Voronoi tessellation classification. The results show that thick, wetting amorphous intergranular films have multiple types of structural order at amorphous-crystalline interfaces and inside the film interior, which in turn is comprised of a transition region and a region that has short-range order identical to that of bulk amorphous phases. However, no fully saturated region exists if the film is too thin. The thickness of the transition region depends on film thickness at low temperatures, but high temperature removes this effect. The structural order in amorphous complexions is also shown to be dependent on interfacing crystal planes at low temperatures but high temperatures remove this behavior. These findings demonstrate that amorphous complexions are structurally distinct from bulk amorphous materials, with an interesting variety of spatial gradients in short-range order within the films.

## Acknowledgements

This research was supported by U.S. Department of Energy, Office of Science, Basic Energy Sciences under Award # DE-SC0014232.

**Figures and Captions**

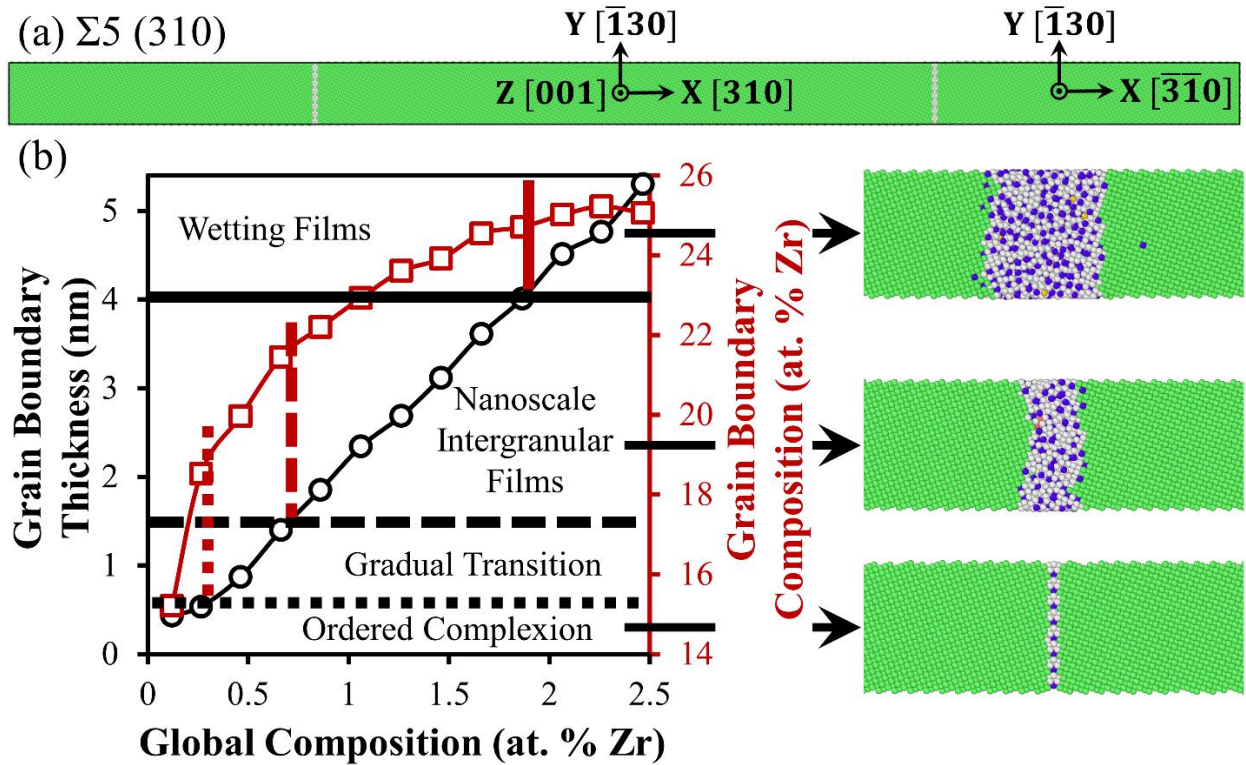

Figure 1. (a) A bicrystal sample containing two Σ5 (310) grain boundaries used to prepare amorphous intergranular films through segregation-induced disordering transitions. (b) Grain boundary thickness and composition for the sample as a function of global composition when doped with Zr at 600 K. The dotted, dashed, and solid lines mark the boundaries between ordered complexion (Type I), gradual transition region, nanoscale intergranular film (complexion Type V), and wetting film (complexion Type VI), with typical equilibrium structures of the three complexion types shown to the right.

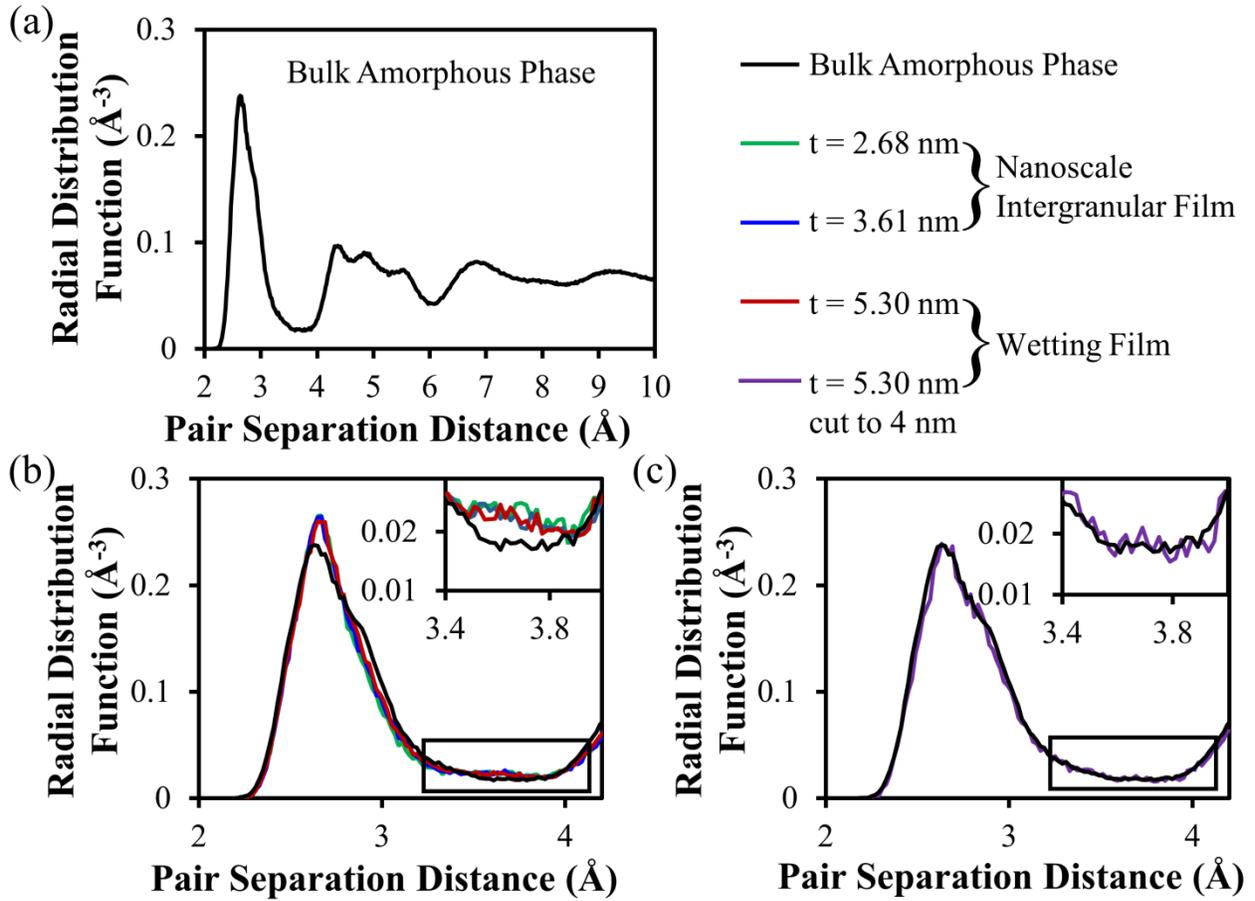

Figure 2. Radial distribution functions of (a) bulk amorphous phase, (b) two nanoscale intergranular films and one wetting films obtained at 600 K, and (c) the 5.30 nm thick wetting film cut to 4 nm thick. Insets to (b) and (c) show the radial distribution function at the valley after the first peak.

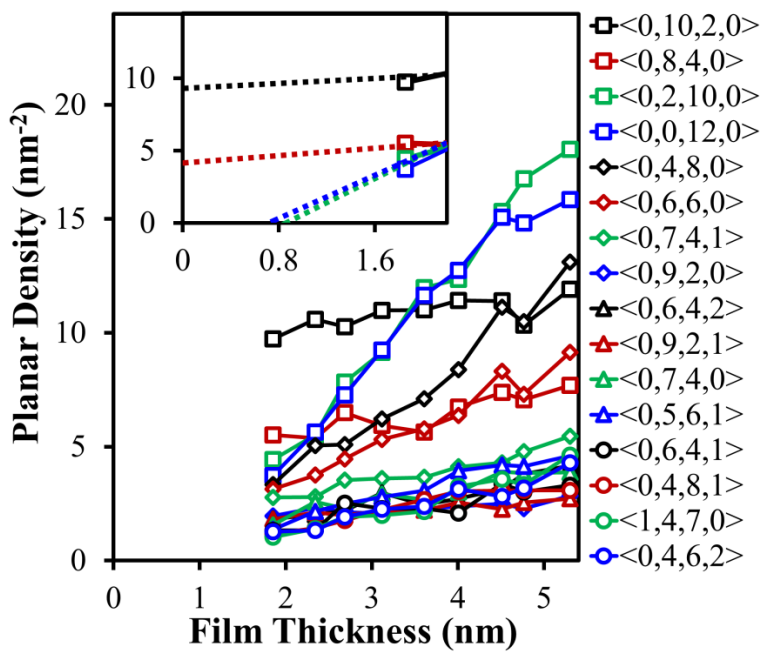

Figure 3. Planar density of Cu atoms associated with a collection of 10 most frequent Voronoi polyhedra in amorphous films transformed from the Σ5 (310) grain boundary at 600 K as a function of film thickness. There are no data points for film thickness between 0 and ~1.7 nm since the amorphous films do not completely cover the grain boundary in this range. The inset to this figure shows the planar density when the film thickness is extrapolated towards zero, where the fitted lines of four polyhedra are shown to intercept with the vertical or horizontal axis, indicating that some polyhedra appear at amorphous-crystalline interfaces and some only in the film interior.

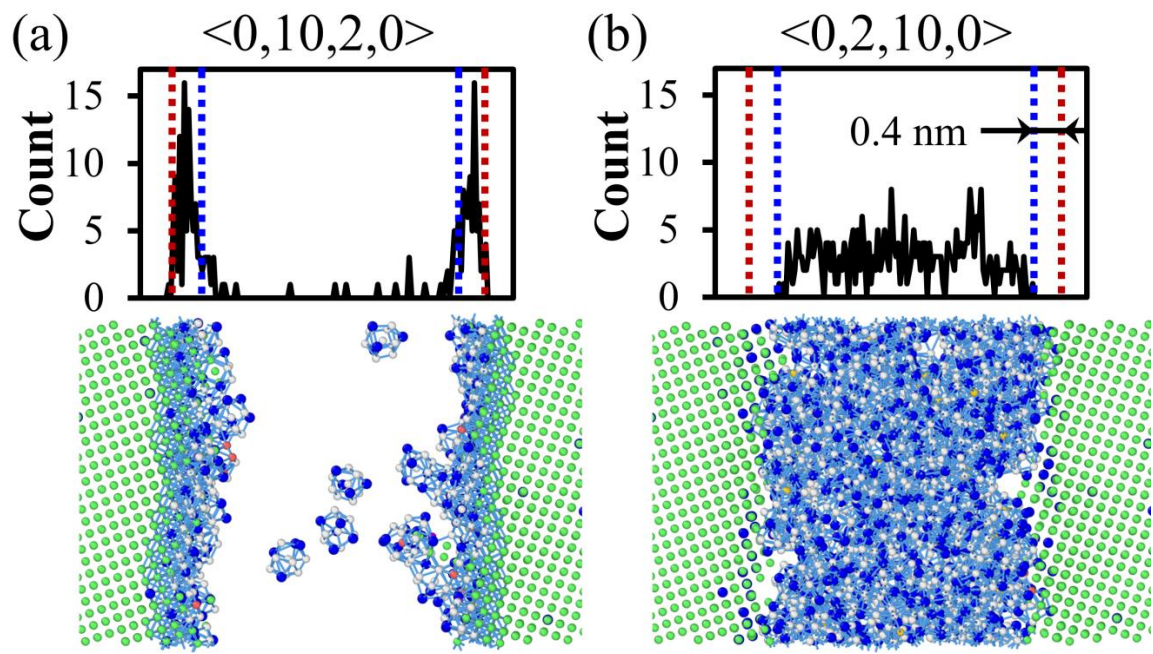

Figure 4. Distribution of (a) <0,10,2,0> and (b) <0,2,10,0> polyhedra along the thickness direction of a 4.51 nm thick amorphous intergranular film transformed from the Σ5 (310) grain boundary at 600 K. The neighbors of each center atom and the coordination polyhedron formed by the neighbors is shown as well. The vertical red and blue dotted lines mark the boundaries of the abutting crystals and of the disordered film interior, respectively.

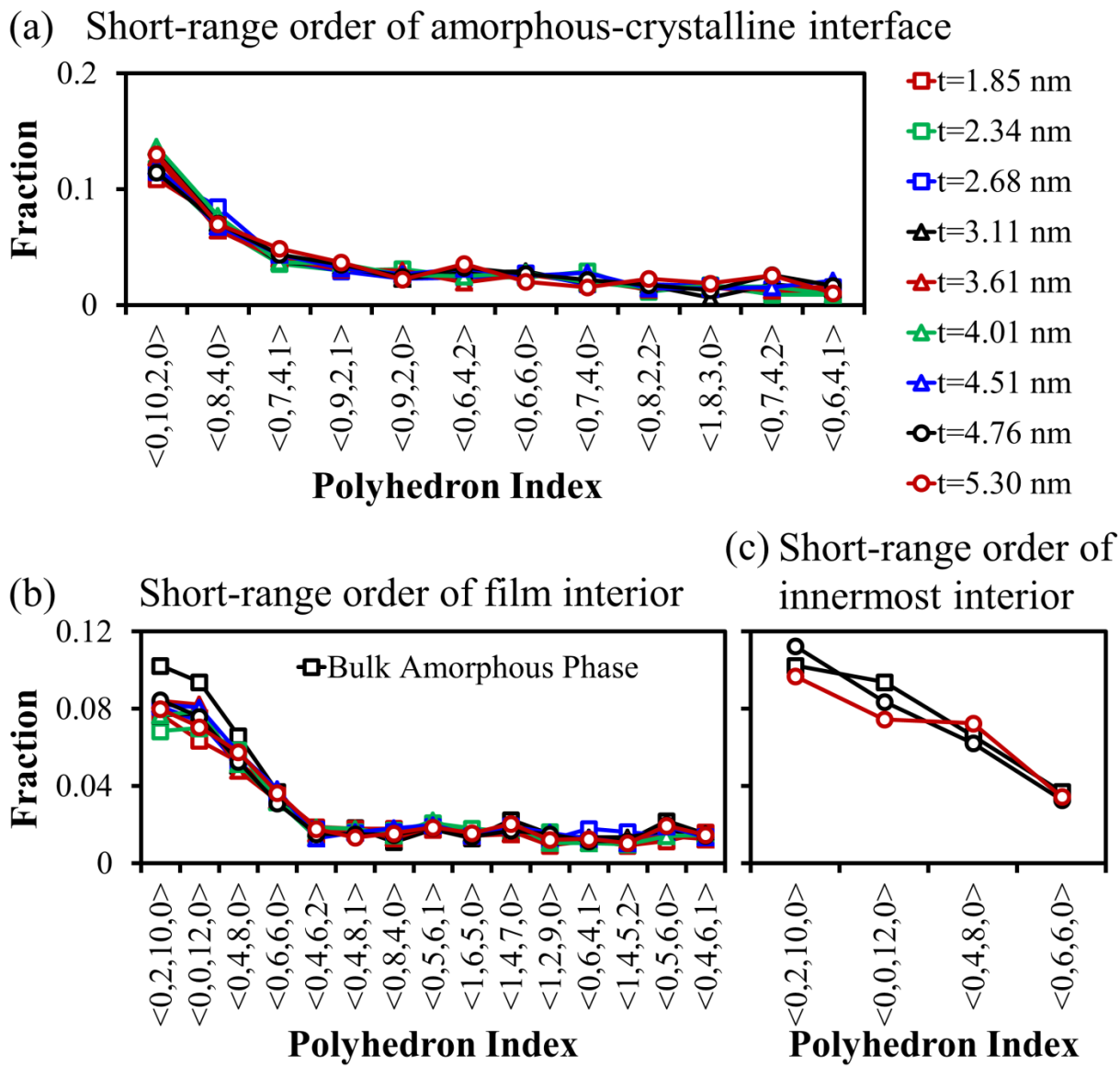

Figure 5. The fraction of atoms associated with a collection of 10 most frequent polyhedra in (a) amorphous-crystalline interfaces and (b) film interior. (c) The fraction of the four dominate polyhedra shown in (b) in the 4.76 and 5.30 nm thick film's innermost interior that is at least 2 nm away from the crystals. The amorphous films are transformed from the Σ5 (310) grain boundary at 600 K. The data for bulk amorphous phase obtained at 600 K is shown in (b) and (c) as well to make a comparison.

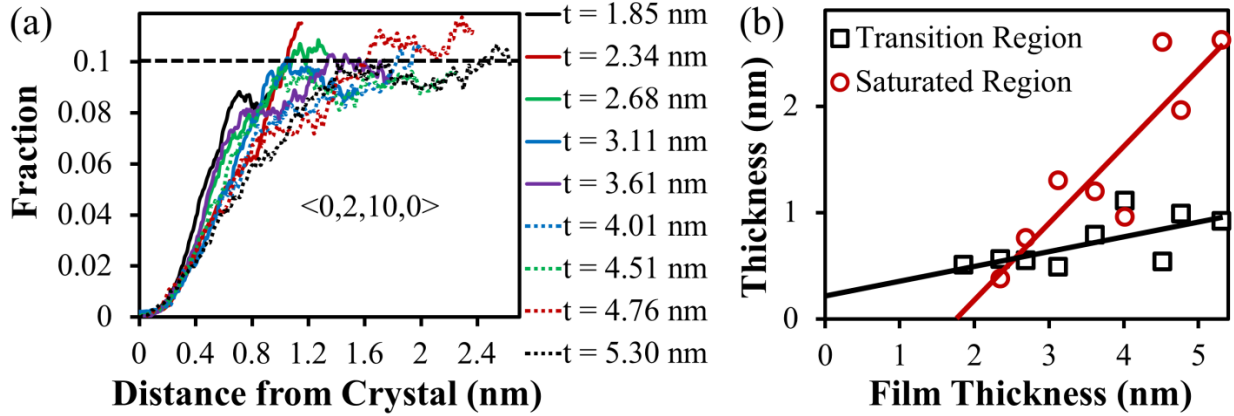

Figure 6. (a) Fraction distribution of <0,2,10,0> polyhedron from film boundary to film center in amorphous complexions transformed from the Σ5 (310) grain boundary at 600 K. (b) The thickness of transition region and saturated region measured based on the <0,2,10,0> fraction distribution. The fraction is considered saturated when reaching a value of 0.09, which is 90% of the <0,2,10,0> polyhedron fraction (shown as the horizontal dashed line in (a)) in bulk amorphous phases.

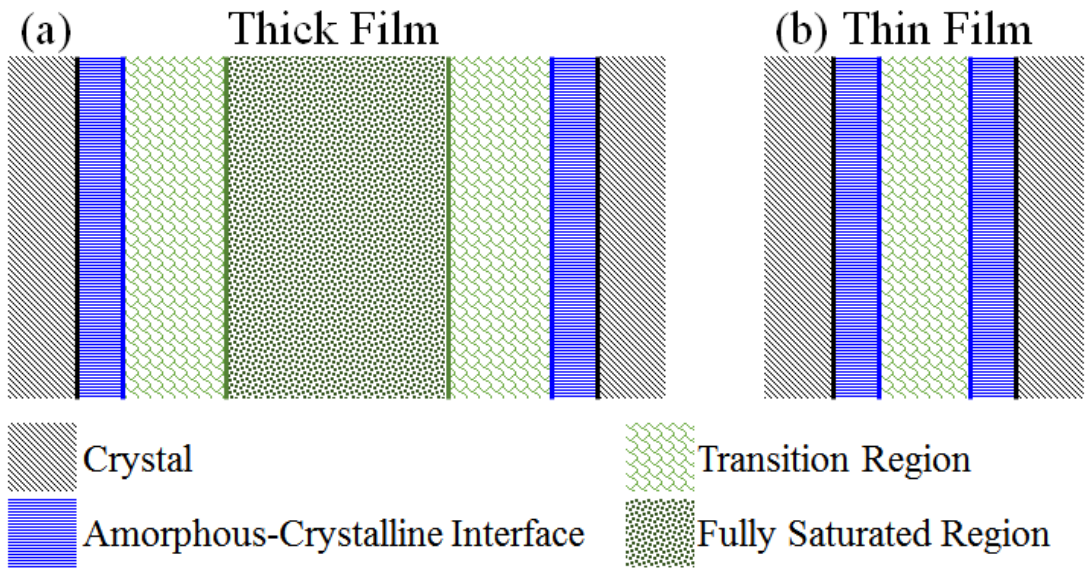

Figure 7. Schematic of different constituent regions for (a) thick and (b) thin amorphous intergranular films.

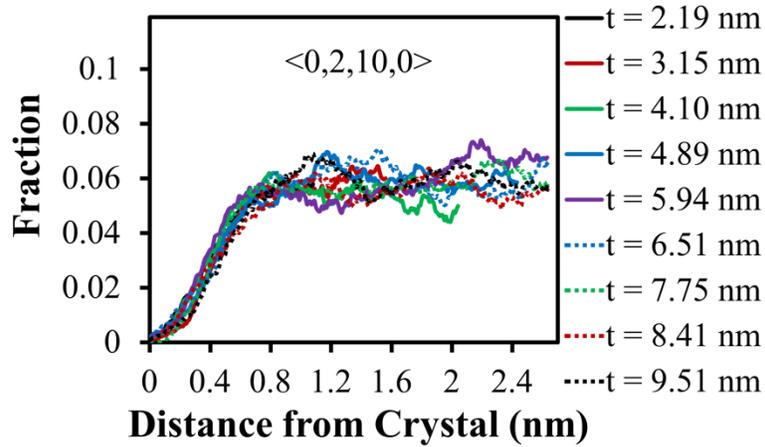

Figure 8. Fraction distribution of <0,2,10,0> polyhedron from film boundary to film center in amorphous complexions transformed from the Σ5 (310) grain boundary at 1200 K. Consistent <0,2,10,0> polyhedron distribution is observed for films of different thickness. The data at distances from film boundary larger than 2.7 nm follows the same plateau and is therefore not shown here.

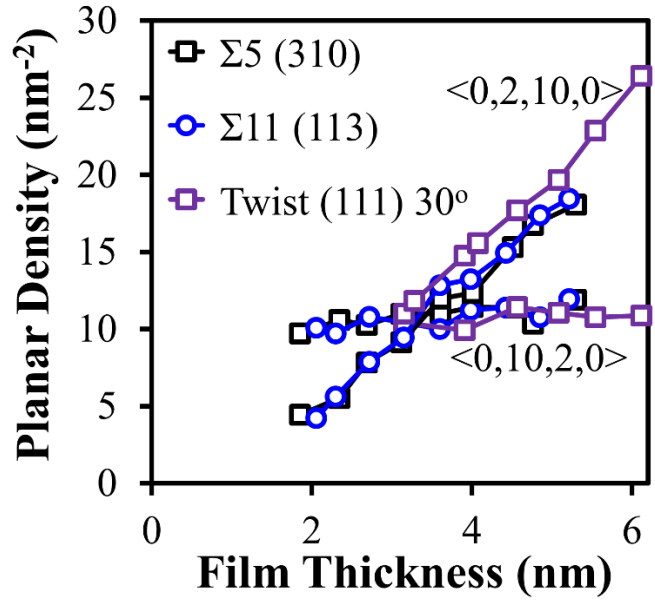

Figure 9. The planar density of atoms associated with <0,2,10,0> and <0,10,2,0> polyhedra in amorphous films transformed from various grain boundaries at 600 K as a function of film thickness.

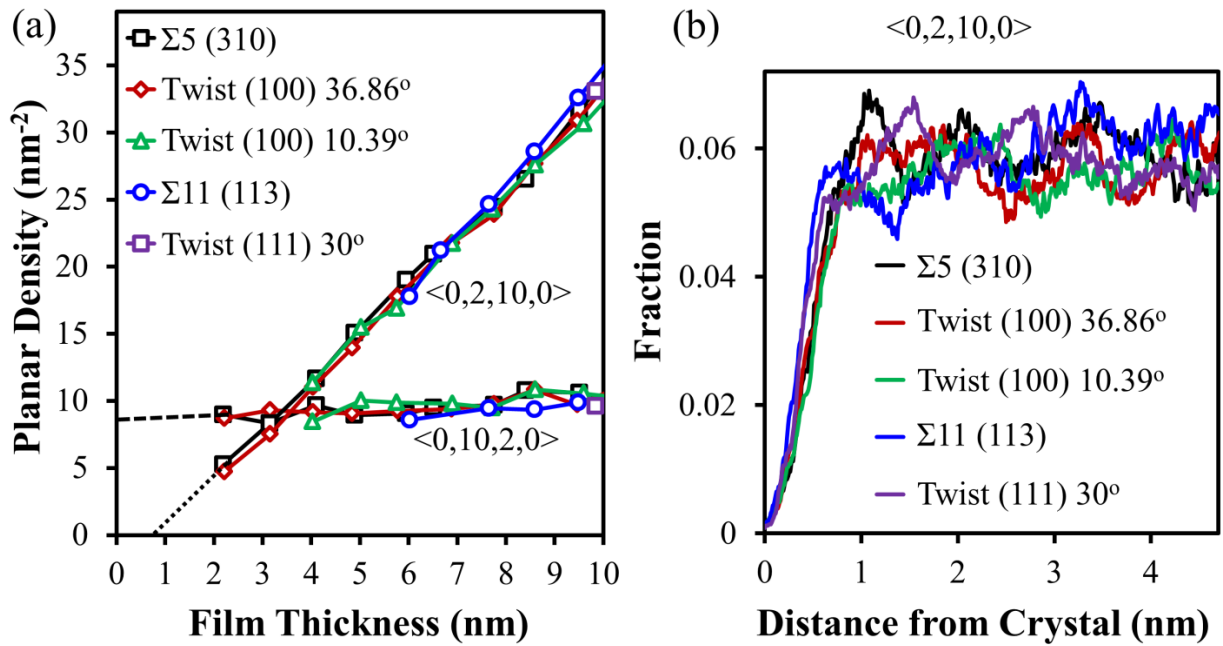

Figure 10. (a) The planar density of atoms associated with <0,2,10,0> and <0,10,2,0> polyhedra in amorphous films transformed from various grain boundaries at 1200 K as a function of film thickness. The dotted and dashed lines show the trend of <0,2,10,0> and <0,10,2,0>, respectively, when reducing the film thickness to zero. (b) The fraction distribution of atoms associated with <0,2,10,0> polyhedron in ~10 nm thick amorphous films transformed from various grain boundaries at 1200 K.